\documentclass[aps,prb,twocolumn,superscriptaddress]{revtex4-2}
\usepackage{graphicx}
\usepackage{amsmath,amssymb}
\usepackage{bm}
\usepackage{color}
\usepackage{hyperref}
\usepackage{siunitx}

\usepackage[normalem]{ulem}
\begin{document}

\title{Casimir forces on suspended thin dielectric membranes covered with graphene}

 \author{Tze Ue Ngai}
\affiliation{Department of Physics, The Hong Kong University of Science and Technology, Clear Water Bay, Kowloon, Hong Kong, China}
\affiliation{William Mong Institute of Nano Science and Technology, The Hong Kong University of Science and Technology, Clear Water Bay, Kowloon, Hong Kong, China}
\affiliation{IAS Center for Quantum Matter, The Hong Kong University of Science and Technology, Clear Water Bay, Kowloon, Hong Kong, China}

\author{Zuocheng Zhang}
\affiliation{Department of Physics, The Hong Kong University of Science and Technology, Clear Water Bay, Kowloon, Hong Kong, China}
\affiliation{William Mong Institute of Nano Science and Technology, The Hong Kong University of Science and Technology, Clear Water Bay, Kowloon, Hong Kong, China}
\affiliation{IAS Center for Quantum Matter, The Hong Kong University of Science and Technology, Clear Water Bay, Kowloon, Hong Kong, China}

\author{Q.~Zhang}
\affiliation{Department of Physics, The Hong Kong University of Science and Technology, Clear Water Bay, Kowloon, Hong Kong, China}
\affiliation{William Mong Institute of Nano Science and Technology, The Hong Kong University of Science and Technology, Clear Water Bay, Kowloon, Hong Kong, China}
\affiliation{IAS Center for Quantum Matter, The Hong Kong University of Science and Technology, Clear Water Bay, Kowloon, Hong Kong, China}

\author{Youssef Jeyar}
\affiliation{Laboratoire Charles Coulomb (L2C), UMR 5221 CNRS-Universit\'{e} de Montpellier, F-34095 Montpellier, France}

\author{Jingwei Li}
\affiliation{Department of Chemical and Biological Engineering, The Hong Kong University of Science and Technology, Clear Water Bay, Kowloon, Hong Kong, China}

\author{Zhengtang Luo}
\affiliation{Department of Chemical and Biological Engineering, The Hong Kong University of Science and Technology, Clear Water Bay, Kowloon, Hong Kong, China}
\affiliation{William Mong Institute of Nano Science and Technology, The Hong Kong University of Science and Technology, Clear Water Bay, Kowloon, Hong Kong, China}

\author{Mauro Antezza}
\affiliation{Laboratoire Charles Coulomb (L2C), UMR 5221 CNRS-Universit\'{e} de Montpellier, F-34095 Montpellier, France}
\affiliation{Institut Universitaire de France, 1 rue Descartes, Paris Cedex 05 F-75231, France}

\author{H.~B.~Chan}
\email{hochan@ust.hk}
\affiliation{Department of Physics, The Hong Kong University of Science and Technology, Clear Water Bay, Kowloon, Hong Kong, China}
\affiliation{William Mong Institute of Nano Science and Technology, The Hong Kong University of Science and Technology, Clear Water Bay, Kowloon, Hong Kong, China}
\affiliation{IAS Center for Quantum Matter, The Hong Kong University of Science and Technology, Clear Water Bay, Kowloon, Hong Kong, China}

\begin{abstract}
We investigate the Casimir force on thin suspended dielectric layers covered with graphene, both theoretically and experimentally. Thermal effects of the Casimir force on graphene occur at separations significantly smaller compared to ordinary metals and dielectrics. At such small separations, the Casimir force is sufficiently strong to be measured experimentally. However, the presence of substrates to support the graphene significantly reduces the enhancement. We show theoretically that replacing the solid substrate beneath graphene with a thin dielectric membrane partially restores this strong thermal dependence. One of the interacting objects is a solid gold plate. The other object is a suspended dielectric membrane made of either 15~nm of silicon oxide or 13~nm silicon nitride, covered with graphene. At a separation of $\sim$500~nm, theoretical predictions yield changes in the Casimir pressure of up to 54\% and 40\% for silicon oxide and silicon nitride, respectively, when the temperature is increased from 10~K to 294~K. Measurements were performed for graphene on a suspended silicon nitride membrane at room temperature, using a gold-coated sphere attached to a cantilever as the force sensor. The measured force gradient is consistent with the theoretical prediction, although the thermal dependence could not be resolved, in part because of uncertainties in the optical properties of silicon nitride. Replacing a bulk substrate with a thin dielectric membrane covered with graphene provides a substantial reduction of the Casimir force, which may facilitate the stable operation of neutral nanomechanical elements at small separations.
\end{abstract}

\maketitle

\section{\label{sec:introduction}INTRODUCTION}

Casimir forces refer to the interactions between electrically neutral bodies that arise due to fluctuations. At zero temperature, the Casimir force originates from quantum fluctuations of the electromagnetic field~\cite{Casimir1948} and the corresponding polarization fluctuations induced in the bodies~\cite{Lifshitz1956}. The Casimir force is practically negligible for objects separated by large distances. However, it increases rapidly with decreasing separation and becomes the dominant interaction between uncharged bodies at submicron separations. Over the past three decades, advances in microfabrication have enabled the development of ultrasensitive transducers, allowing the Casimir force to be measured with high sensitivity~\cite{Bordag2009, Dalvit2011, Decca2003, Decca2007, Chan2008, Munday2009, deMan2009, Torricelli2010, Sushkov2011, Chang2012, Laurent2012, Pires2021, Zhang2024}.

At non-zero temperatures, thermal fluctuations also contribute to the Casimir force. The theoretical formulation of the finite-temperature Casimir force was established within the framework of fluctuation-dissipation theory~\cite{Lifshitz1956}, in which the electromagnetic response is evaluated at discrete imaginary Matsubara frequencies $\xi_n = 2\pi n k_B T/\hbar$ (with $n = 0, 1, 2, \ldots$). The characteristic length scale governing the onset of thermal contributions is set by the thermal wavelength $\lambda_T = \hbar c / (k_B T)$, which is approximately 7.8~$\mu$m at room temperature ($T$ = 294 K). For conventional 3D bulk conductors and dielectrics, thermal corrections become appreciable only at separations comparable to or larger than $\lambda_T$. At such large separations, however, the Casimir force becomes weak and can be masked by residual interactions. The thermal Casimir force has been experimentally demonstrated at separations up to 7~$\mu$m~\cite{Sushkov2011}, where residual electrostatic forces had to be subtracted from the measured force.

Graphene fundamentally alters the thermal Casimir interaction due to its unique electronic structure, which exhibits linear energy dispersion near the Dirac points~\cite{Gomez-Santos2009,Drosdoff2010,Svetovoy2011,Sernelius2011,Bordag2012,Klimchitskaya2013,Klimchitskaya2014b,Bordag2015,Abbas2017,Bimonte2017,Jeyar2023,Jeyar2023a,Jeyar2025}. The low\mbox{-}energy excitations in graphene are characterized by a Fermi velocity approximately 300 times smaller than the speed of light. Consequently, thermal effects on the Casimir force emerge at separations significantly smaller than those required for conventional metals and dielectrics. At these short separations, the Casimir force remains sufficiently strong to be experimentally accessible. Early theoretical studies of the Casimir interaction in graphene considered isolated graphene layers, without supporting substrates~\cite{Gomez-Santos2009}. Subsequent theoretical analyses showed that thermal effects persist when graphene is supported by a substrate, although their magnitude is reduced~\cite{Klimchitskaya2014b,Bimonte2017}. The Casimir force was first measured between a gold sphere and graphene on a silicon wafer covered with 285~nm of silicon oxide~\cite{Banishev2013}. More recently, thermal effects have been experimentally observed at separations as small as 250~nm by using a much thicker silicon oxide layer~\cite{Liu2021b,Liu2021a}.

While suspended graphene yields more pronounced thermal effects, experimental studies have utilized graphene on solid substrates due to practical constraints. First, it is challenging to fabricate a graphene layer that is suspended, large, and clean simultaneously. While suspended structures are rather common in various studies of graphene, the suspended parts are typically only a few micrometers large for monolayer graphene~\cite{Bunch2007,Lee2008,Takahashi2018}. Other applications requiring large areas often use multilayer graphene~\cite{Barton2011,Li2015,Pezone2022,Wan2026}. Measurement of Casimir forces requires monolayer graphene samples with an interaction area of up to $50~\mu\text{m}^2$ and minimal organic contamination. The cleaning steps involving aggressive solvents or annealing can easily damage suspended graphene, particularly for large suspended areas. Second, suspended graphene is susceptible to significant deformation under external forces, which reduces the effective separation and thereby complicates the interpretation of Casimir force measurements.

In this article, we investigate, both theoretically and experimentally, the Casimir force between a solid gold surface and a thin, suspended dielectric layer covered with graphene. One of the goals is to address the difficulties in fabricating suspended graphene. Initially, the dielectric layer is chosen to be silicon oxide for its well-known optical properties. The Casimir pressure exhibits an enhanced temperature dependence for a 15~nm-thick suspended silicon oxide layer covered with graphene compared with a structure in which the suspended layer is replaced by a bulk substrate. Then we replace the silicon oxide membrane with a silicon nitride membrane that is robust enough to survive the fabrication steps. Due to incomplete knowledge of the optical properties of the silicon nitride, several assumptions are required to calculate the Casimir pressure. We find that the Casimir pressure also exhibits an enhanced temperature dependence for a 13~nm-thick silicon nitride membrane covered with graphene compared with a structure in which the membrane is replaced by a bulk substrate. In the experiments, we fabricated a sample consisting of suspended silicon nitride covered by graphene. While the measured force gradient is consistent with theoretical calculations, uncertainties in the properties of the graphene and silicon nitride layers preclude a demonstration of the thermal effects.

We further find that the force gradient scaled by the radius of the sphere is significantly reduced compared with previous measurements using much thicker interacting bodies. The Casimir force becomes the dominant interaction between neutral mechanical components when their separations are reduced to sub-micrometer levels. Reducing the Casimir force or changing the sign of the force can alleviate undesirable stiction between mechanical components. However, for planar surfaces at full thermal equilibrium, repulsive Casimir forces can only be achieved with the introduction of a liquid with suitable dielectric properties into the gap between the interacting bodies~\cite{Munday2009}. Our approach provides an alternative to significantly diminish the attractive Casimir force by reducing the thickness of one of the interacting bodies to $\sim$10~nm.

In Sec.~\ref{sec:calculations}, we present the theoretical calculations of the Casimir pressure. In Sec.~\ref{sec:measurement}, we describe the sample fabrication and experimental setup, and present and discuss the experimental results.
%


\section{\label{sec:calculations}THEORETICAL CALCULATIONS OF THE CASIMIR PRESSURE}

\subsection{Theoretical framework}
\label{sec:theory}

\begin{figure}[htbp]
\centering
\includegraphics[width=0.85\columnwidth]{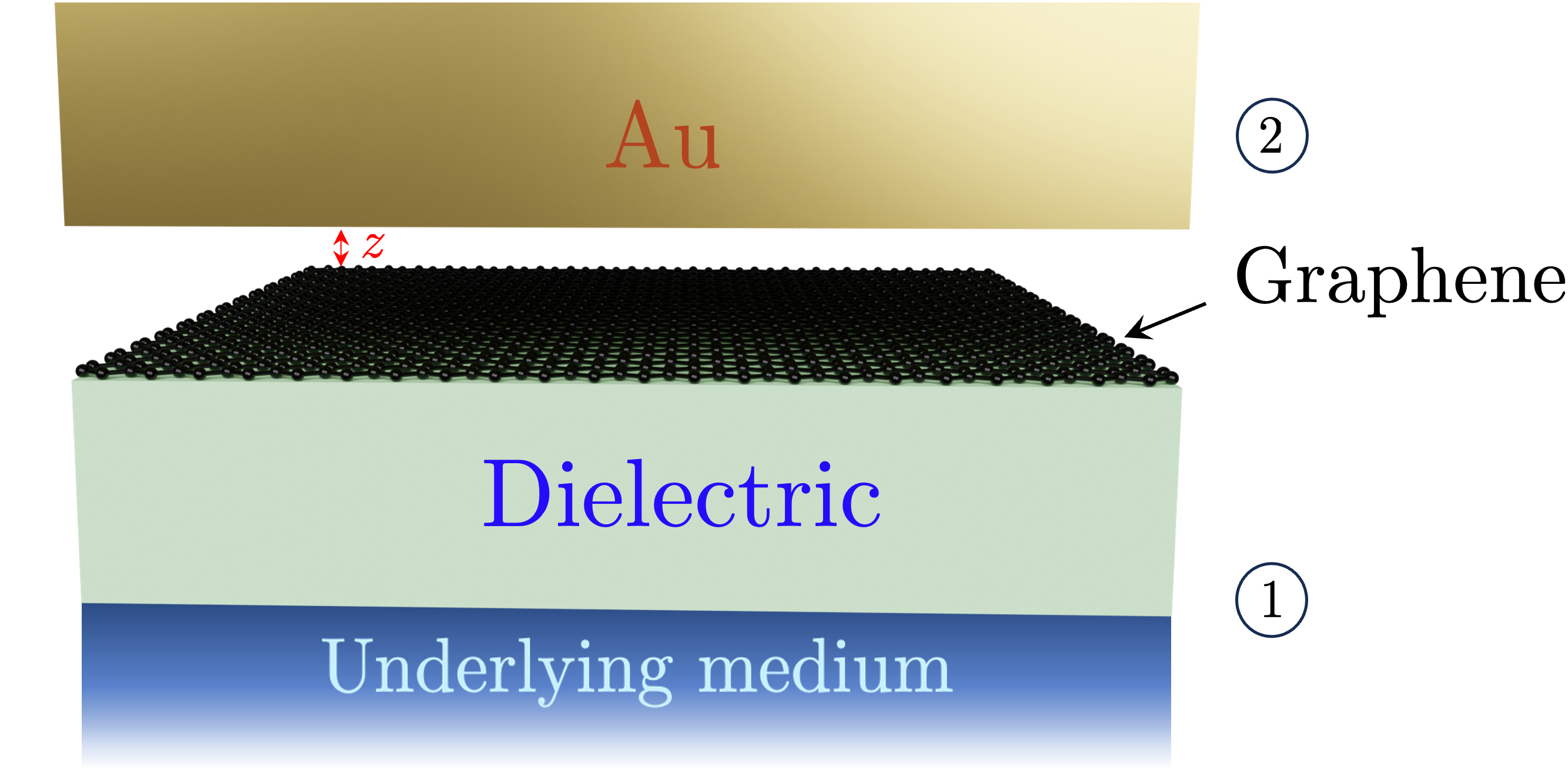}
\caption{Schematic representation of the structure considered in the theoretical calculations.}
\label{fig:structure}
\end{figure}

The physical system depicted in Fig.~\ref{fig:structure} consists of a graphene-covered dielectric structure (body~1) facing a gold half-space (body~2). The two bodies are separated by a distance $z$ and maintained at a temperature $T$. The dielectric structure can be either a semi-infinite medium, a finite thickness layer, or a finite-thickness layer supported by a substrate. \\

The Casimir pressure between the two bodies is calculated using the Lifshitz scattering formalism~\cite{Abbas2017} and is given by
\begin{equation}
P =
\frac{k_{\mathrm B}T}{\pi}
\sum_{n=0}^{\infty}{}'
\int_0^\infty dQ\,Qq
\sum_{p}
\left[
\frac{e^{2qz}}
{R_p^{(1)}R_p^{(2)}}
-1
\right]^{-1},
\label{eq:casimir_pressure}
\end{equation}
where the prime indicates that the $n=0$ term is divided by two,
$\xi_n=2\pi n k_{\mathrm B}T/\hbar$ are the Matsubara frequencies, and
\begin{equation}
q=\sqrt{\left(\frac{\xi_n}{c}\right)^2+Q^2}.
\end{equation}
Here, $Q$ is the magnitude of the wave vector parallel to the interfaces, $p$ denotes the polarization, and $R_p^{(i)}(Q,i\xi_n,T)$ denotes the reflection coefficient of the $i$th interacting structure, calculated exactly following the expressions given in the Supplemental Material of Ref.~\cite{Abbas2017}. The positive sign of $P$ corresponds to an attractive interaction between the two bodies.
For the sphere-plane geometry considered experimentally, the Casimir force gradient is related to the Casimir pressure between parallel
plates through the proximity force approximation (PFA)~\cite{Bimonte2018,Chan2008,Jeyar2024},

\begin{equation}
F' = 2\pi R P,
\label{eq:force_gradient}
\end{equation}
where $R$ is the radius of the sphere.

The graphene response is described by its frequency-dependent surface conductivity, which, on the imaginary frequency axis, is written as the sum of intraband and interband contributions,
\begin{equation}
\sigma(i\xi_n)
=
\sigma_{\mathrm{intra}}(i\xi_n)
+
\sigma_{\mathrm{inter}}(i\xi_n),
\end{equation}
where
\begin{equation}
\sigma_{\mathrm{intra}}(i\xi_n)
=
\frac{8\sigma_0 k_{\mathrm B}T}
{\pi(\hbar\xi_n+\hbar\Gamma)}
\ln\left[
2\cosh\left(\frac{\mu}{2k_{\mathrm B}T}\right)
\right],
\end{equation}
and
\begin{equation}
\sigma_{\mathrm{inter}}(i\xi_n)
=
\frac{4\sigma_0\hbar\xi_n}{\pi}
\int_0^\infty
\frac{G(x)}
{(\hbar\xi_n)^2+4x^2}\,dx.
\end{equation}
Here, $\sigma_0=e^2/(4\hbar)$, $e$ is the electron charge, $\mu$ is the graphene chemical potential, and $\Gamma=10^{13}~\mathrm{rad/s}$ is the relaxation rate. The function $G(x)$ is defined as
\begin{equation}
G(x)=
\frac{\sinh(x/k_{\mathrm B}T)}
{\cosh(\mu/k_{\mathrm B}T)+\cosh(x/k_{\mathrm B}T)}.
\end{equation}

The dielectric function of the gold used in this work is described by the Drude model,
\begin{equation}
\epsilon_{\mathrm{Au}}(i\xi_n)
=
1+
\frac{\omega_p^2}
{\xi_n(\xi_n+\gamma)},
\label{eq:drude}
\end{equation}
with $\hbar\omega_p=9~\mathrm{eV}$ and
$\hbar\gamma=35~\mathrm{meV}$.

\subsection{Temperature dependence for silicon oxide membranes}
\label{sec:sio2}

\begin{figure}[htbp]
\centering
\includegraphics[width=\columnwidth]{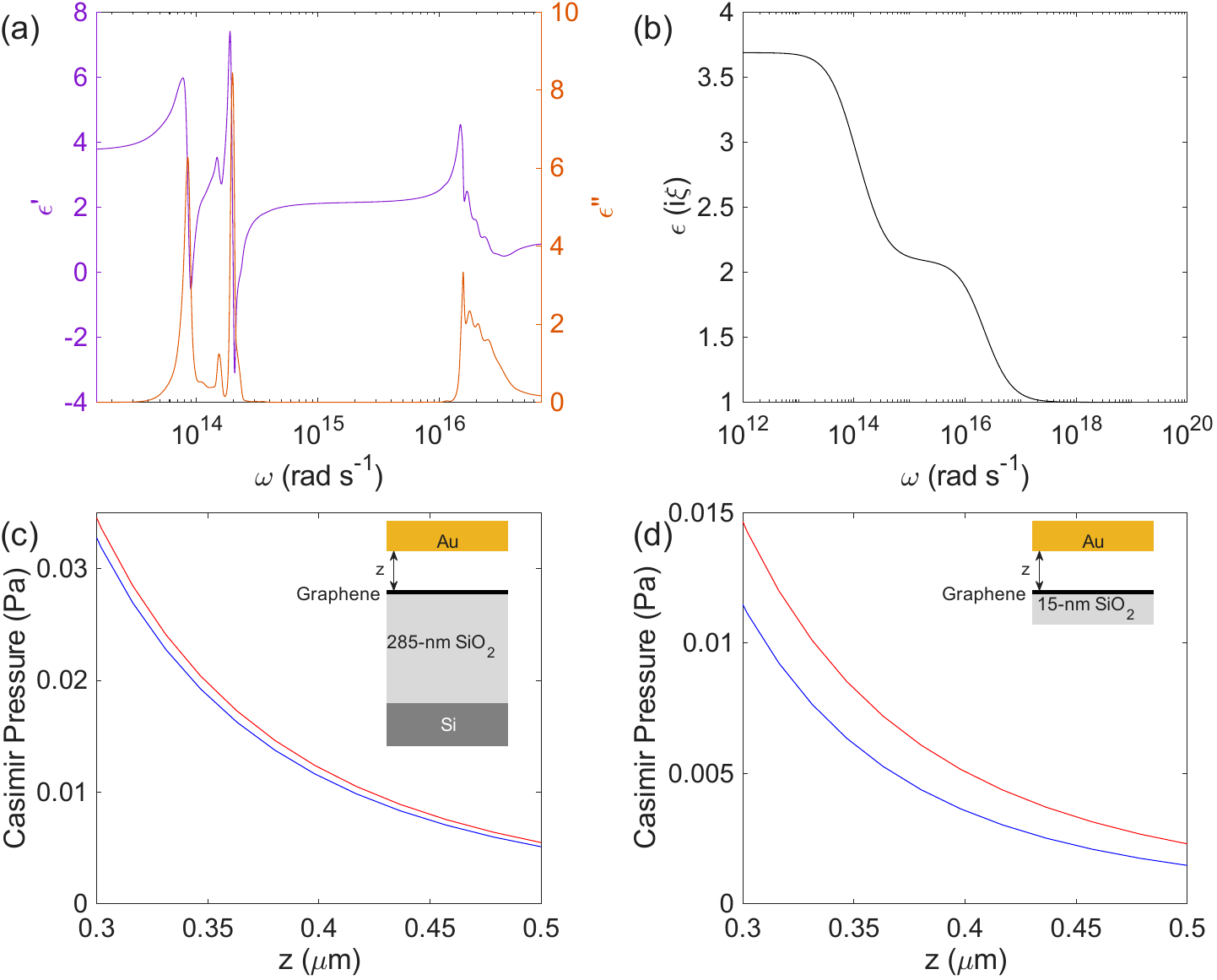}
\caption{(a) Real (purple) and imaginary (brown) parts of the dielectric function of silicon oxide as a function of frequency~\cite{Palik1997}. (b) The dielectric permittivity for imaginary frequencies calculated using Kramers-Kronig relations. (c) Inset (not to scale): a solid gold planar object is located at separation $z$ from a 285~nm-thick silicon oxide on a silicon substrate, covered with monolayer graphene with chemical potential of 165~meV. Main figure: The Casimir pressure as a function of $z$ for temperature of 10~K (blue) and 294~K (red). (d) The lower object is changed to monolayer graphene on a suspended 15~nm-thick silicon oxide. Thermal effects are significantly enhanced.}
\label{fig:1}
\end{figure}

The dielectric function of silicon oxide has been measured over a broad range of wavelengths and is tabulated in data books~\cite{Palik1997}. Figure~\ref{fig:1}(a) shows the real part $\epsilon'(\omega)$ and imaginary part $\epsilon''(\omega)$ of the dielectric function of silicon oxide as a function of frequency. The dielectric response of silicon oxide at imaginary frequencies is obtained through the Kramers-Kronig relation $\epsilon(i\xi) = 1 + 2\pi^{-1}\int_0^\infty \omega\epsilon''(\omega)/(\omega^2+\xi^2)\,d\omega$, relying on the data of Ref.~\cite{Palik1997} and is plotted in Fig.~\ref{fig:1}(b). The resulting $\epsilon(i\xi)$ is a smooth function of frequency. As $\xi$ approaches zero, the value of $\epsilon(i\xi)$ is consistent with the low-frequency limit of $\epsilon'(\omega)$. For configurations involving a silicon substrate, the dielectric response of silicon is included using the model and parameters of Ref.~\cite{Wang2021}.
\color{black}

Figure~\ref{fig:1}(c) shows the Casimir pressure calculated between a bulk gold slab and graphene (chemical potential $\mu = 165$~meV) placed on a 285~nm-thick layer of silicon oxide on a silicon substrate, over a distance range $z$ of 300~nm to 500~nm. At $z = 479$~nm, the Casimir pressure at $T = 294$~K exceeds that at $T = 10$~K by only 7\%. Thermal effects are largely masked by the presence of the thick silicon oxide and silicon substrate. If the substrate is replaced by a 15~nm thick layer of silicon oxide that is suspended without the silicon wafer, the change in the Casimir pressure between the two temperatures is increased to $\sim$54\% [Fig.~\ref{fig:1}(d)], which, in principle, is readily measurable in experiments. Deformations of the membrane in response to the force are ignored in the calculations. In experiments, the amount of deformation depends on the size of the suspended area and the thickness of the oxide membrane.

While thin silicon oxide membranes can be fabricated, the material is under compressive stress. Instead of staying flat, the suspended oxide membrane is extremely fragile and tends to crumble. Transfer of graphene onto a target substrate requires removal of the backing substrate by solvents and additional cleaning steps to remove organic residues. Transferring graphene onto suspended oxide membranes often leads to breakage. We have not been successful in fabricating a sample that is intact, clean, and large enough for Casimir force measurements at the same time. On the other hand, silicon nitride membranes are widely used in transmission electron microscopy and optomechanical systems. They are durable and remain flat due to the tensile stress in the material. Furthermore, for stoichiometric silicon nitride membranes, the tensile stress is high so that mechanical deformations in response to the Casimir force can be minimized.

\subsection{Temperature dependence for silicon nitride membranes}

\label{sec:sin}

\begin{figure}[htbp]
\centering
\includegraphics[width=\columnwidth]{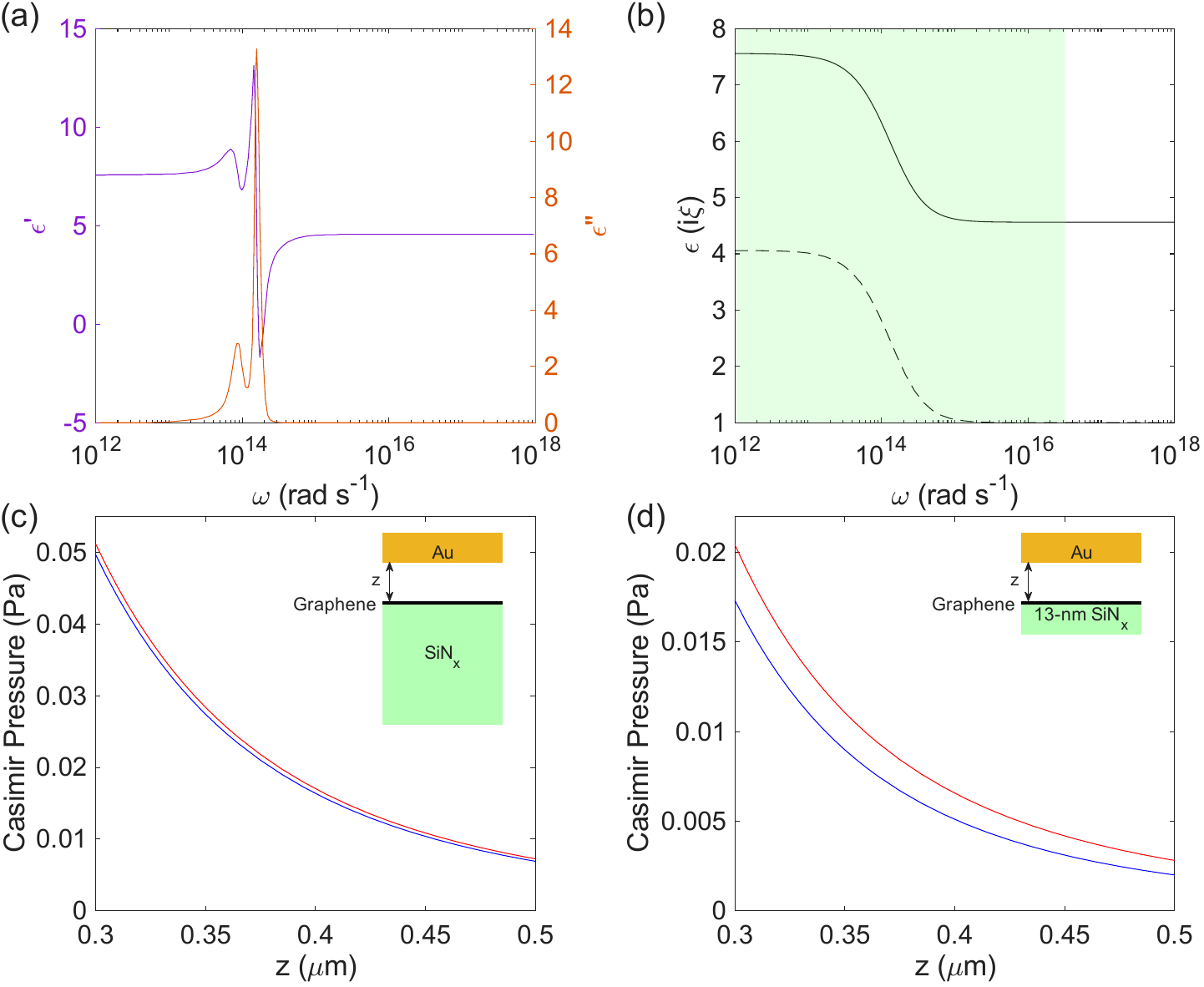}
\caption{(a) Real (purple) and imaginary (brown) parts of the dielectric function of silicon nitride as a function of frequency~\cite{Cataldo2012}. (b) The dashed curve shows the dielectric permittivity for imaginary frequencies calculated with Kramers-Kronig relations. The solid curve is obtained by shifting the results by 3.562 so that $\epsilon(i\xi)$ at $\xi = 0$ agrees with $\epsilon'(\omega=0)$ in (a). Calculations of the Casimir pressure are performed in the green-shaded frequency region and are terminated before $10^{16}$~rad~s$^{-1}$. (c) Inset (not to scale): a solid gold planar object is located at separation $z$ from solid silicon nitride covered with monolayer graphene with chemical potential of 165~meV. Main figure: The Casimir pressure as a function of $z$ for temperature of 10~K (blue) and 294~K (red). (d) The lower object is changed to monolayer graphene on a suspended 13~nm-thick silicon nitride. Thermal effects are significantly enhanced.}
\label{fig:2}
\end{figure}

While the mechanical properties of silicon nitride membranes are more desirable compared to silicon oxide membranes, they suffer from a main disadvantage in Casimir force measurements because the optical properties are not as thoroughly characterized. The optical properties may also depend on the composition and the deposition conditions. Figure~\ref{fig:2}(a) shows the dielectric functions of silicon nitride measured between $\omega$ ranging from $10^{12}$~rad~s$^{-1}$ to $10^{18}$~rad~s$^{-1}$ from Ref.~\cite{Cataldo2012}. Even though the range of $\omega$ exceeds that of the silicon oxide in Fig.~\ref{fig:1}(a), difficulties arise in the calculation of the Casimir pressure. We obtain the response at imaginary frequencies $\epsilon(i\xi)$ from $\epsilon''(\omega)$ through the Kramers-Kronig relation. As $\xi\to0$, the resulting value of $\epsilon(i\xi)$ in Fig.~\ref{fig:2}(b) falls short of the known static value $\epsilon(0)\approx8$, and does not match the low-frequency behavior of $\epsilon'(\omega)$ in Fig.~\ref{fig:2}(a). The discrepancy arises because the Kramers-Kronig relation requires integrating $\epsilon''(\omega)$ over the full oscillator spectrum, while the measured window in Fig.~\ref{fig:2}(a) only resolves the infrared absorption band of the Si–N vibrational modes, peaked  around $1.6\times10^{14}$~rad~s$^{-1}$: as in the case of silicon oxide [Fig.~\ref{fig:1}(a)], each absorption band  in $\epsilon''(\omega)$ produces, in $\epsilon(i\xi)$, a step down to a lower plateau, so that the missing static weight indicates the presence of at least one further resonance beyond the upper edge of the measured range. Since its position and lineshape are not experimentally accessible, we account for its contribution by shifting $\epsilon(i\xi)$ up by a constant value of 3.562, so as to recover the correct static value $\epsilon(0)$. This constant-shift model remains valid only below the frequency of the missing resonance, that is, before the onset of the corresponding plateau, which is precisely where the Matsubara sum entering the Casimir pressure calculation naturally converges, as indicated by the shaded region in Fig.~\ref{fig:2}(b).

Figure~\ref{fig:2}(c) shows the Casimir pressure calculated between a gold slab and graphene on top of solid SiN$_x$. The pressure at $T = 294$~K is larger than that at $T = 10$~K by only 5\% at 495~nm. When the thick silicon nitride is replaced by a suspended silicon nitride membrane with a thickness of 13~nm [Fig.~\ref{fig:2}(d)], the difference is significantly enhanced to 40\%. Our calculations of the Casimir pressure on graphene placed on both silicon oxide and silicon nitride support the notion that the use of suspended membranes to support the graphene can enhance the changes in the Casimir pressure due to thermal effects.

\section{\label{sec:measurement}MEASUREMENT OF THE CASIMIR FORCE GRADIENT ON SILICON NITRIDE MEMBRANES COVERED WITH GRAPHENE}

\subsection{Sample fabrication}
\label{sec:fabrication}

\begin{figure}[htbp]
\centering
\includegraphics[width=\columnwidth]{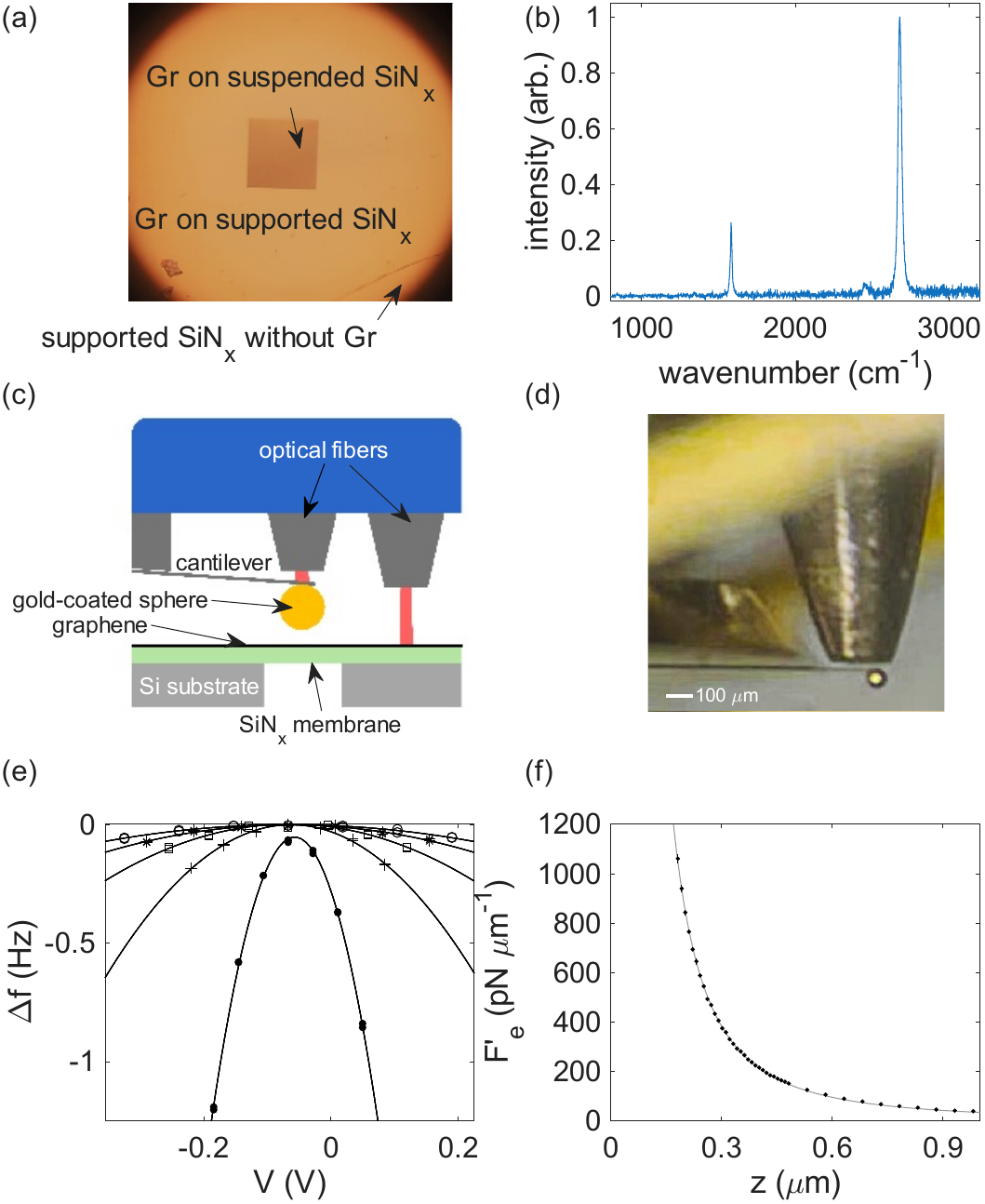}
\caption{(a) Photograph of suspended SiN$_x$ covered with graphene (Gr) used in the force measurement. The suspended SiN$_x$ is $\sim$13~nm thick and 100~$\mu$m by 100~$\mu$m in area. The graphene coverage ends at the lower right corner, exposing the SiN$_x$. (b) Raman spectrum of the graphene on SiN$_x$. (c) Sideview sketch of the experimental setup. A gold-coated sphere is attached to a cantilever. Graphene on suspended SiN$_x$ is placed at the bottom. Vibrations of the cantilever are measured using a fiber interferometer to infer the force gradient. A second fiber interferometer eliminates long term drifts in the separation $z$ between the sphere and the graphene. (d) Photograph of a typical cantilever, sphere, and the optical fiber. The scale bar measures 100~$\mu$m. (e) Measured frequency shift as a function of voltage applied to the graphene at distance $z$ of 1800~nm (hollow circles), 1400~nm (star), 1000~nm (square), 600~nm (cross), and 200~nm (solid circles). Solid lines are parabolic fits. (f) Electrostatic force gradient used for calibration of constants $C$ and $z_0$. The circles and line are measurement and fit using Eq.~(\ref{eq:electrostatic}), respectively.}
\label{fig:3}
\end{figure}

Our samples are made using the conventional procedure of transferring monolayer graphene grown on copper onto the target substrate. The target substrate is either a solid piece of silicon with 285~nm of silicon oxide or a suspended silicon nitride membrane supported by a silicon frame. Figure~\ref{fig:3}(a) shows the suspended 100~$\mu$m by 100~$\mu$m silicon nitride membrane covered by graphene used in force measurement. The bare chip without graphene was purchased from Norcada (TA301Z, suspended silicon nitride thickness ranging from 8~nm to 15~nm). It consists of a silicon frame that supports the suspended non-stoichiometric silicon nitride (SiN$_x$) membrane at its periphery.

The graphene on copper was purchased from a commercial vendor (Grolltex). Poly(methyl methacrylate) (PMMA) was spin-coated onto the copper with graphene. The PMMA served as a support layer for the graphene as the copper was removed using electrochemical delamination~\cite{Wang2011,Gao2012}. After displacing the electrolyte with deionized water, the target substrate was then submerged into the water underneath the floating graphene/PMMA. The substrate was slowly lifted out of the water so that the graphene made contact with the target area, with the PMMA layer on top. After the sample was dried, it was heated to 150\,$^\circ$C to melt the PMMA. Finally, the PMMA was removed using acetone and the sample was rinsed in isopropanol followed by deionized water. Gold electrical contacts to the graphene were made at the edge of each sample using shadow masks.

We measured the Raman spectrum of the graphene on suspended SiN$_x$ membrane used in the force measurement. The microscope focused the beam at the membrane center with a spot radius of 5~$\mu$m. In Fig.~\ref{fig:3}(b), the symmetric 2D peak (2679~cm$^{-1}$) confirmed the presence of monolayer graphene on the membrane. The small ratio of the intensity of D (1341~cm$^{-1}$) to G peak (1585~cm$^{-1}$) suggests that the graphene layer is largely free of defects. Similar Raman spectra were also measured for the graphene on solid silicon oxide.

To estimate the chemical potential of the graphene used in the force experiment, we measured the conductance of another sample consisting of graphene on 285~nm silicon oxide, with the $p$-doped silicon substrate serving as a back gate. By identifying the minimum in conductance as a function of the backgate voltage, we estimate that the chemical potential is 0.165~eV. In this measurement the 285~nm thick silicon oxide gives good optical contrast for identifying the locations of good graphene coverage and for patterning the electrical contacts. The optical contrast for SiN$_x$ with and without graphene, on the other hand, is much weaker compared to 285~nm of silicon oxide. As a result, we did not attempt to make a sample in which the graphene conductance can be modulated by a backgate.

\subsection{Experimental Setup and calibration}
\label{sec:setup}

In the force measurement, the other surface to confine the electromagnetic modes is a hollow glass sphere coated with 150~nm of gold. Two different spheres were used. The radii of sphere~1 and sphere~2 are 38.7~$\mu$m and 43.4~$\mu$m respectively. As shown in Figs.~\ref{fig:3}(c) and \ref{fig:3}(d), each sphere was attached to a gold-coated silicon cantilever (224~$\mu$m by 29~$\mu$m by 3.4~$\mu$m) that served as a resonant detector of the force gradient. The graphene on either solid silicon oxide or a suspended SiN$_x$ membrane was placed at the bottom, sitting on top of piezoelectric positioning stages that controlled the distance to the sphere at the top. A fiber interferometer using 1550~nm light was used to eliminate long term drifts in the separation between the two interacting surfaces, a scheme adopted from previous experiments~\cite{Decca2003}. The entire setup was enclosed in a bell jar maintained at a pressure below $10^{-6}$~torr by a turbomolecular pump backed by a dry roughing pump.

Vibrations of the cantilever were excited by a piezoelectric actuator and detected using a second fiber interferometer using 1310~nm light. The resonance frequencies $f_0$  of the cantilevers with sphere 1 and sphere 2 are $\sim$40963 Hz and $\sim$28795 Hz respectively, with quality factors of 12053 and 9068 respectively. A phase-locked loop tracked the shifts in the resonance frequency of the cantilever as the separation $z$ between the sphere and the graphene was changed by the piezoelectric stage. Vibration amplitudes were kept small so that nonlinear effects were negligible. Since the spatial gradient of the total force $F$ on the sphere is much smaller than the spring constant $k$ of the cantilever ($\sim$1~N~m$^{-1}$), the resonance frequency shift $\Delta f$ of the cantilever is proportional to $F'(z)$:
\begin{equation}
\ \Delta f = CF'(z),
\label{eq:freqshift}
\end{equation}
where $C = -f_0/2k$. The separation $z$ between the graphene and the bottom of the sphere is given by $z = z_0 - z_{\text{piezo}} $, where $z_0$ is the initial separation and $z_{\text{piezo}}$ is the extension of the piezoelectric positioner. Bending of the cantilever produces negligible changes in $z$.

A dc voltage $V$ was applied to the graphene while the sphere remained electrically grounded. The electrostatic force between the two surfaces is given by
\begin{equation}
F_e = 2\pi\epsilon_0(V-V_0)^2\sum\limits_{n=1}^{\infty}\frac{\coth{\alpha}-n\coth{n\alpha}}{\sinh(n\alpha)},
\label{eq:electrostatic}
\end{equation}
where $V_0$ is the residual voltage arising from different work functions of the two surfaces and $\alpha = \cosh^{\text{-1}}(1+z/R)$. Figure~\ref{fig:3}(e) shows the dependence of the measured $\Delta f$ as a function of $V$ for the graphene on SiN$_x$ membrane at five fixed values of $z$. $V_0$ was found to be constant to within measurement uncertainty as $z$ was varied. Calibration of $z_0$ and the proportionality constant $C$ was done by fitting the measured $\Delta f$ to Eqs.~(\ref{eq:freqshift}) and~(\ref{eq:electrostatic}) for six values of $V-V_0$ ($\pm$100~mV, $\pm$140~mV, and $\pm$170~mV) after subtracting the contribution of the Casimir force gradient. Figure~\ref{fig:3}(f) shows the fit of the electrostatic force gradient at $V=V_0+170$ mV. In the measurement of the Casimir force gradient, $V$ was set equal to $V_0$ and the force gradient $F'_C$ was deduced from $\Delta f$ using the values of $C$ and $z_0$ obtained from the electrostatic fits. Sphere~1 was used in the first experiment to measure the force gradient on solid silicon oxide covered by graphene. A different sphere (sphere~2) was used to measure the force gradient on suspended nitride membrane covered by graphene after the first sphere was accidentally damaged. For the two measurements, the fitted values of $C$ are $(-3.17\pm0.13)\times10^3$ s kg$^\text{-1}$ and $(-2.184\pm0.027)\times10^3$ s kg$^\text{-1}$ respectively.

\subsection{Measurement results}
\label{sec:results}

\begin{figure}[htbp]
\centering
\includegraphics[width=\columnwidth]{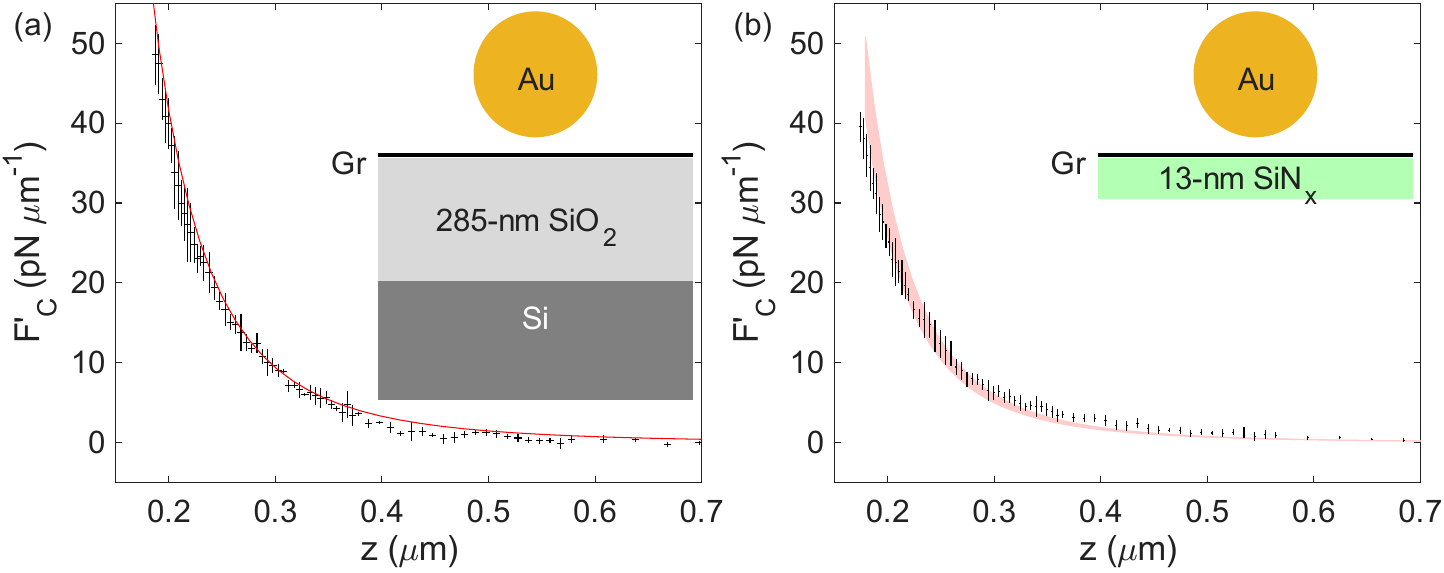}
\caption{(a) Inset (not to scale): Gold sphere~1 with radius 38.7~$\mu$m is placed on top of 285~nm-thick silicon oxide on a silicon substrate, covered with monolayer graphene. Main figure: Red line represents calculated Casimir force gradient as a function of distance $z$. Measured results are shown in black. (b) Inset: Gold sphere~2 with radius 43.4~$\mu$m is placed on top of 13~nm thick suspended SiN$_x$ covered with monolayer graphene. Main figure: The light-red band shows the range of calculated Casimir force gradient obtained by varying the SiN$_x$ thickness from 10~nm to 16~nm. Measured results are shown in black.}
\label{fig:4}
\end{figure}

Figure~\ref{fig:4}(a) shows the measured force gradient between gold sphere~1 and graphene on 285~nm of SiO$_2$ deposited on a silicon wafer after the residual voltage was compensated. The solid line represents the Casimir force gradient calculated by theory, using $\mu = 165$~meV. Roughness of the gold surface ($\sim$3.8~nm rms) and the graphene ($\sim$0.64~nm rms) are small compared to the distance $z$. It gives small corrections to the calculated Casimir force gradient~\cite{Bordag2009}. The measured results are consistent with theoretical calculations at 294~K. However, measurement uncertainties are too large to indicate difference from the force gradient at 10~K to show thermal effects.

In Fig.~\ref{fig:4}(b), the measurement was repeated for gold sphere~2 and another sample where the graphene is placed on a suspended silicon nitride membrane. As expected, by suspending the thin nitride membrane, the force gradient becomes smaller than that on thick silicon oxide in Fig.~\ref{fig:4}(a). The calculated force gradient is shown as a light-red band. Discrepancies between calculations and experiments arise from a number of sources. First, as described in earlier sections, the dielectric function of silicon nitride is not known over a sufficiently large frequency range for the Kramers-Kronig integration to give consistent values with the dielectric function at low, real frequencies. The optical properties of the SiN$_x$ used in our experiment may also differ from Ref.~\cite{Cataldo2012} due to different composition and deposition conditions. Second, there are uncertainties in the average thickness of the SiN$_x$ and non-uniformities across the sample. The membrane vendor specifies that within a membrane, thickness variations can be as large as 1~nm. We used an atomic force microscope to measure the thickness of a piece of silicon nitride broken away from another chip that ultimately landed on a silicon substrate and obtained a value of 13~nm. The thickness of the light-red band in Fig.~\ref{fig:4}(b) accounts for thickness variations of about $\pm$3~nm between different suspended nitride samples. Third, while the chemical potential can be estimated for graphene on silicon oxide, much less has been reported for graphene on silicon nitride. While previous experiments~\cite{Liu2021b,Liu2021a} have used the shift of the G peak in the Raman spectra to estimate the chemical potential for graphene on oxide, data for graphene on silicon nitride is not available. For the calculations shown in Fig.~\ref{fig:4}(b), we use a chemical potential for the calculated curve that is the same as the graphene on silicon oxide. Lastly, while we adopt fabrication steps to minimize the amount of organic residues from transferring the graphene onto the target substrates, we cannot completely rule out the presence of organic residues that might trap charges and alter the force.

One of the motivations of using dielectric membranes to support the graphene is to minimize the deformation of the suspended layers in response to Casimir or electrostatic forces. The deformation of a membrane depends on a number of factors including the Young's modulus of the material, the thickness and area of the membrane, the pre-tension in the absence of external force, and load distribution~\cite{Wan2003}. To estimate the deformation, we use an external force of 200~pN that exceeds all electrostatic and Casimir forces in the measurement and assume that the force is uniformly distributed on a 30~$\mu$m radius circle at the center of the membrane. The resultant deformation is less than 0.3~nm, justifying our assumption that the SiN$_x$ remains flat throughout the experiment.

By replacing the bulk substrate with a 13~nm-thick silicon nitride membrane and putting a monolayer of graphene on top, we achieved, to our knowledge, the smallest Casimir force gradient measured so far in the sphere-plane configuration. We compare the force gradients measured in experiments at a distance of 250~nm. Applying the proximity force approximation, the ratio of the force gradient between a gold sphere and a planar test body to the radius of the sphere multiplied by $2\pi$ is identical to the pressure between a flat gold surface and the test body. After scaling by the radius of the gold sphere, the force gradient for the graphene on the suspended silicon nitride membrane is a factor of $\sim$5 smaller than previous experiments using gold-coated surfaces~\cite{Chang2012,Decca2007}. Compared to prior measurements of solid dielectric layers covered with graphene, the force gradient we measured is $\sim$40\% smaller~\cite{Liu2021a}. Reduction of the Casimir force can facilitate the stable operation of neutral nanomechanical elements in tight proximity.

\section{\label{sec:summary}SUMMARY}

In summary, we study the Casimir force on thin suspended dielectric layers covered by graphene, both in calculations and experiments. Our calculations show that by replacing a solid substrate underneath the graphene by a thin dielectric membrane, thermal dependence of the Casimir pressure is significantly enhanced compared to graphene on solid substrates. The Casimir force gradient was measured between a gold-coated sphere and graphene on solid silicon oxide or on a suspended SiN$_x$ membrane. Measurements were consistent with theory but uncertainties in the properties of the silicon nitride membrane and the graphene preclude comparison between measurement and calculations at a level that can distinguish thermal effects. Future efforts can improve the experiment by directly measuring the optical properties of the silicon nitride layer used in the experiment, over a wide frequency range. The thickness of the membrane and the uncertainty in the thickness can be reduced by performing the fabrication in house. Alternatively, the graphene can be supported by other dielectrics such as few-layer hexagonal boron nitride to further reduce the influence of the substrate on the thermal dependence of the Casimir force.

\begin{acknowledgments}
The work described in this paper was supported by a grant ``CAT'' from the ANR/RGC Joint Research Scheme sponsored by the French National Research Agency (ANR) and the Research Grants Council (RGC) of the Hong Kong Special Administrative Region, China (Project No.~A-HKUST604/20).
\end{acknowledgments}

\bibliography{ngai}

\end{document}